\documentclass[10pt,conference,review]{IEEEtran}
\IEEEoverridecommandlockouts
\usepackage{cite}
\usepackage{amsmath,amssymb,amsfonts}
\usepackage{algorithmic}
\usepackage{graphicx}
\usepackage{textcomp}
\usepackage[table]{xcolor}
\usepackage{booktabs}
\usepackage{enumitem}
\usepackage{booktabs}
\usepackage{multirow}
\usepackage{graphicx}
\usepackage{caption}
\usepackage[table]{xcolor}
\usepackage{amsmath,amsfonts}
\usepackage{graphicx}
\usepackage{textcomp}
\usepackage{xcolor}
\usepackage[hidelinks]{hyperref}
\usepackage[most]{tcolorbox}

\newcommand{\dataset}{CLI-Tool-Bench}

\definecolor{darkgreen}{rgb}{0,0.5,0}

\def\BibTeX{{\rm B\kern-.05em{\sc i\kern-.025em b}\kern-.08em
    T\kern-.1667em\lower.7ex\hbox{E}\kern-.125emX}}

\tcbset{
    myfindingbox/.style={ % 定义一个名为 myfindingbox 的样式
        colback=gray!10, % 背景颜色：10% 的灰色。你可以调整这个值来改变深浅。
        colframe=white, % 边框颜色：白色（实际上是隐藏边框，因为背景色会覆盖它）
        boxrule=0pt, % 边框粗细：0pt 表示没有可见边框
        arc=1mm, % 圆角半径：4mm，你可以根据需要调整这个值
        outer arc=1mm, % 外部圆角半径，与 arc 保持一致
        left=1mm, right=1mm, top=1mm, bottom=1mm, % 内边距，让文本与边框有一定距离
    }
}

\begin{document}

% \author{\IEEEauthorblockN{Anonymous Author(s)}
% \IEEEauthorblockA{\textit{Affiliation} \\
% City, Country \\
% email address or ORCID}
% }

\title{Evaluating LLM-Based 0-to-1 Software Generation in End-to-End CLI Tool Scenarios}
% \title{An Empirical Study on End-to-End LLM-based 0-to-1 Software Generation}
\author{Ruida Hu$^{1}$, Xinchen Wang$^{1}$, Chao Peng$^{2}$, Cuiyun Gao$^{1,\dagger}$, David Lo$^{3}$%
  \thanks{$^{1}$ Harbin Institute of Technology, Shenzhen, China (email: \{200111107, 200111115\}@stu.hit.edu.cn, gaocuiyun@hit.edu.cn)}%
  \thanks{$^{2}$ Independent Researcher, China (email: chao.peng@acm.org)}%
  \thanks{$^{3}$ Singapore Management University, Singapore (email: davidlo@smu.edu.sg)}%
  \thanks{$^{\dagger}$ Corresponding author.}%
}
\maketitle

% \section{Abstract}
\begin{abstract}
The evolution of Large Language Models (LLMs) has catalyzed a paradigm shift towards intent-driven software development, where autonomous agents are expected to %architect
design and deliver complete, runnable software systems %entirely 
from scratch. However, existing benchmarks fail to adequately assess this 0-to-1 generation capability due to two fundamental limitations. First, they rely on predefined structural scaffolds, which reduces the task to mere file-filling. Second, they depend on rigid white-box unit testing, which forces generated code to conform to specific internal implementations rather than validating end-to-end user-centric behavior.

To bridge this gap, we introduce \dataset, a novel, structure-agnostic benchmark designed to evaluate the ground-up generation of Command-Line Interface (CLI) tools. Powered by an automated black-box differential testing framework, the benchmark comprises 94 high-quality, real-world repositories spanning diverse programming languages and %complexities
complexity levels. For each task, agents are provided with an empty workspace, forcing them to autonomously handle repository planning and dependencies. %We evaluate the generated software by executing it in isolated sandboxes and comparing its system-level side effects and terminal outputs against human-written oracles using a rigorous multi-tiered equivalence metric.
We evaluate the generated software by executing it in isolated sandboxes. The system-level side effects and terminal outputs are then compared against human-written oracles using a rigorous multi-tiered equivalence metric.

Extensive evaluation of seven state-of-the-art LLMs reveals that the top-tier models achieve a maximum overall success rate of only 43.8\%, highlighting that 0-to-1 software generation remains a highly challenging frontier. Furthermore, we discover that agents exhibit a strong tendency to generate monolithic code structures, and that higher token consumption does not necessarily yield better task performance.
\end{abstract}

\section{Introduction}
The advent of Large Language Models (LLMs) has catalyzed a paradigm shift in automated software engineering~\cite{DBLP:journals/corr/abs-2310-13976, DBLP:journals/corr/abs-2403-14274, llmsurvey, llmsurvey2}. Moving beyond simple code completion, recent advancements have spurred the development of autonomous LLM-based agents~\cite{Repairagent, coder, DBLP:journals/corr/abs-2409-02977, DBLP:journals/corr/abs-2310-16340} capable of tackling complex programming tasks. This rapid progress is driving the industry toward the era of ``Vibe Coding'' or intent-driven development~\cite{ray2025review, maes2025gotchas, meske2025vibe}. In this new paradigm, users only need to express their high-level requirements in natural language. Autonomous agents are then expected to %take
interpret these requirements and generate a complete, runnable software repository %entirely
from scratch.

To measure the capabilities of these agents, robust evaluation benchmarks are indispensable. However, existing benchmarks fall short of evaluating the %true potential
capabilities of agents %in
for real-world software creation. Traditional benchmarks, such as HumanEval~\cite{DBLP:journals/corr/abs-2107-03374} and MBPP~\cite{DBLP:journals/corr/abs-2108-07732}, are confined to function-level generation. While recent efforts like SWE-bench~\cite{DBLP:conf/iclr/JimenezYWYPPN24} have elevated the evaluation to the repository level, they predominantly focus on software maintenance tasks rather than %the creation of software
software creation from scratch. More recently, %attempts 
benchmarks like NL2Repo-Bench~\cite{DBLP:journals/corr/abs-2512-12730} have explored zero-to-one software generation. However, their evaluation methodologies reveal critical limitations that hinder the accurate assessment of modern LLM agents. Specifically, we identify two major challenges in the current evaluation landscape:

\textbf{Challenge 1: Reliance on predefined repository structures.} Evaluating the true 0-to-1 software generation capabilities of LLM agents remains an open challenge. Even in generation-focused benchmarks, the evaluation heavily relies on a fixed, predefined repository structure~\cite{DBLP:journals/corr/abs-2512-12730}. These benchmarks typically provide agents with pre-built file skeletons and directory scaffolds, thereby reducing the complex task of software generation to mere code-filling. This structure-dependent paradigm fundamentally bypasses %a critical step 
the repository structure planning in software creation.
%: repository structure planning. 
In real-world software development, developers must %autonomously
decide how to organize directories, modularize files, and manage dependencies. By constraining agents to predefined structures, current benchmarks fail to assess whether LLMs can independently plan and construct a coherent repository.

\textbf{Challenge 2: Absence of end-to-end black-box testing.} Furthermore, existing evaluations heavily rely on white-box unit testing~\cite{DBLP:conf/iclr/ZhaoJLCCGR25,DBLP:journals/corr/abs-2512-12730,DBLP:conf/iclr/JimenezYWYPPN24,DBLP:journals/corr/abs-2107-03374}. These tests are tightly coupled with the internal implementation details of the software, forcing the generated code to conform to specific function signatures or class definitions. This approach severely %stifles
limits the structural autonomy and creativity of LLMs. More importantly, from a user-centric perspective, software utilities like Command-Line Interface (CLI) tools are %always
typically consumed as black boxes. Users care about whether the tool correctly parses command-line arguments, produces the expected terminal outputs, and 
%executes the correct system-level side effects (e.g., modifying the file system).
produces the expected system-level side effects (e.g., file system modifications). Consequently, the reliance on rigid white-box testing fails to provide %a
realistic and end-to-end validation of these critical external behaviors.

To bridge these gaps, we introduce \dataset, a novel, structure-agnostic benchmark specifically designed to evaluate the 0-to-1 generation of CLI tools. We select CLI tools as our evaluation target because they %are
represent fundamental software utilities. Their externally observable behaviors, including parsing standard inputs and producing terminal outputs or file system changes, make them %ideal candidates
well-suited for rigorous black-box testing without requiring internal code inspection.
We curate a high-quality dataset of 94 real-world CLI repositories across three programming languages (i.e., Python, JavaScript, Go) and various complexity levels. To overcome the reliance on predefined structures, %the 
each agent is provided only with a natural language requirement and an empty workspace, forcing complete autonomy over internal structures and dependencies. We propose a novel evaluation method based on black-box differential testing. Specifically, the generated tool is executed in an isolated sandbox, and its terminal outputs alongside system-level side effects (e.g., file system changes) are rigorously compared against a human-written oracle. To ensure fair assessment without penalizing architectural diversity, we propose a multi-tiered equivalence metric encompassing Execution Reliability, Side-Effect Consistency, and Behavioral Equivalence.
%(Exact, Fuzzy, and Semantic Match).

In summary, the contributions of this paper are as follows:
\begin{itemize}
    \item We introduce \dataset, the first structure-agnostic benchmark for evaluating end-to-end software generation from scratch, which enforces true structural autonomy without reliance on predefined scaffolds.
    \item We propose an automated black-box differential testing pipeline that evaluates generated repositories in isolated sandboxes, rigorously verifying execution reliability, %system-level side effects, 
    side-effect consistency, and behavioral equivalence.
    \item We conduct an extensive evaluation of seven state-of-the-art LLMs and agent frameworks on 94 real-world tasks. %Our results reveal that top models cap at an overall success rate of 43.8\%, underscoring the immense difficulty of the task.
    Our results reveal that the best-performing models achieve a maximum overall success rate of only 43.78\%, underscoring the significant difficulty of this task. We also uncover critical behavioral patterns, such as %agents' 
    the strong preferences of agents for monolithic designs and the prevalence of costly, unproductive debugging cycles.
\end{itemize}

\section{Construction and Evaluation Pipeline}
\begin{figure*}[hbt]
	\centering
	\includegraphics[width=0.9\textwidth]{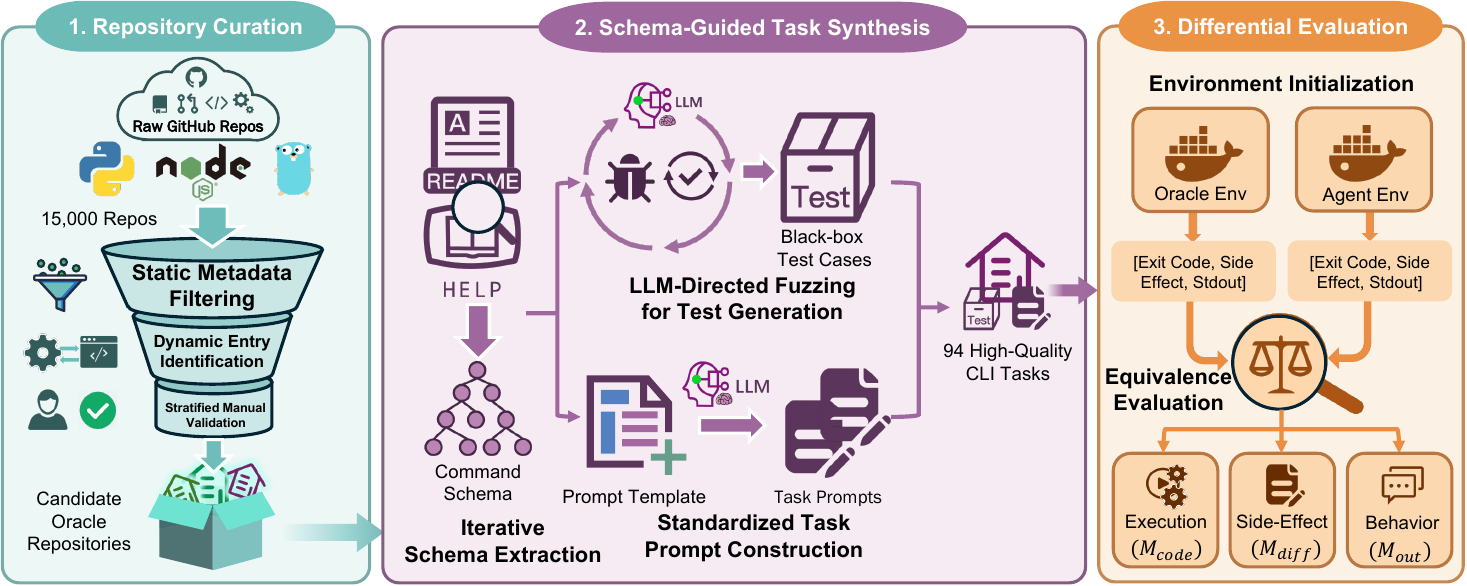}
    % \vspace{-1em}
    \caption{Overview of the \dataset~framework. 
    % The pipeline comprises three core modules: \textbf{(1) Repository Curation} to mine and validate high-quality CLI oracles; \textbf{(2) Task Synthesis} to automatically generate standardized prompts and test cases via LLM-directed fuzzing; and \textbf{(3) Differential Evaluation} to assess the agent-generated tools against the oracles in isolated sandboxes based on execution, behavioral, and side-effect equivalence.
    }
\label{fig:framework}
% \vspace{-1em}
\end{figure*}

To evaluate 0-to-1 software generation, we propose an automated pipeline (Figure~\ref{fig:framework}) comprising three core modules: (1) Repository Curation, (2) Schema-Guided Task Synthesis, and (3) Black-Box Differential Evaluation.

\subsection{Repository curation}
To construct a representative dataset, we target three mainstream programming languages widely adopted for CLI development: Python, JavaScript, and Golang~\cite{li2024multilingual}. Through a multi-stage filtering process, we yield a final curated set of 94 high-quality oracle repositories.

\subsubsection{Static metadata filtering} We initially retrieve candidate repositories from GitHub using specific criteria: 
(1) stars $>$10; (2) primary language proportion $>$60\%; (3) presence of CLI-related keywords; (4) open-source license; and (5) the existence of language-specific build configuration files (e.g., ``\texttt{setup.py}'', ``\texttt{package.json}'', or ``\texttt{go.mod}'').
%(1) Stars $>$10; (2) Primary language purity $>$60\%; (3) Presence of CLI-related keywords; (4) Open-source license; and (5) Existence of language-specific build configuration files (e.g., ``\texttt{setup.py}'', ``\texttt{package.json}'', or ``\texttt{go.mod}'').

\subsubsection{Dynamic entry identification} However, not all retrieved repositories %yield
provide functional standalone CLI tools. To rigorously select valid repositories, we install each candidate globally in an isolated environment. %The core challenge %here 
A key challenge is to automatically identify the exact command name used to invoke the installed CLI, which frequently differs from the repository name. To address this, our pipeline monitors the system's binary directory during %the installation phase
installation to capture all newly generated %executables
executable files. We then compare the repository name against these captured files using a heuristic matching algorithm that prioritizes exact matches, acronyms, and string similarities.
%(prioritizing exact matches, acronyms, and string similarities).
Once the most likely executable file is identified, %we validate it 
it is validated by running the ``\texttt{--help}'' command. 
%If this execution successfully returns standard usage documentation without crashing, it confirms we have located a functional CLI entry point.
If this command successfully returns standard usage information without errors, the repository is considered to contain a functional CLI entry point.

\subsubsection{Stratified manual validation} Repositories that pass this dynamic verification are further subjected to a stratified manual validation by two software engineering experts.
%This step ensures %they execute without underlying environmental errors and rigorously filters out tasks with implicit dependencies (e.g., non-standard local server environments) to guarantee 100\% unambiguous specification. This rigorous filtering process ensures the high quality of our curated oracle repositories.
This step ensures that the repositories can run without environmental errors. It also filters out tasks with implicit dependencies like non-standard local server environments to ensure fully unambiguous specifications. This rigorous filtering process ensures the high quality of the curated oracle repositories.

\subsection{Schema-guided task synthesis}
Generating comprehensive test cases at scale is a critical challenge. We address this by employing an LLM to extract a structured command schema from each curated oracle repository, which subsequently guides the automated generation of evaluation prompts and test suites.

\subsubsection{Iterative schema extraction}
For each oracle repository $R_{oracle}$, the LLM iteratively parses its \texttt{README} documentation and ``\texttt{--help}'' outputs. It explores nested subcommands level by level to extract a hierarchical metadata schema. 
%This schema essentially defines the entire external interface of the CLI, specifying the available command names, subcommands, accepted parameters (and their data types), and strict execution constraints (such as mutually exclusive flags or required arguments).
This schema defines the external interface of the CLI, including available command names, subcommands, accepted parameters and their data types, and strict execution constraints such as mutually exclusive flags or required arguments.

\subsubsection{LLM-directed fuzzing for test generation}
Based on the extracted schema, we implement LLM-directed fuzzing to systematically cover the CLI's capabilities. We first %unroll 
expand the schema into a set of distinct \textbf{Command patterns}. Each pattern represents a specific way to use the CLI,  %(e.g., 
such as a unique subcommand paired with a valid combination of flags. 

For each pattern, the LLM utilizes a predefined testing framework to generate diverse test instances covering common usage, boundary conditions, and error handling. The core of this process is an automated execution-feedback loop. %the 
Each generated test command is directly executed against the human-written oracle $R_{oracle}$. Crucially, the validity of these test cases is %guaranteed 
determined by the oracle itself. Any command that successfully executes on the ground-truth repository without errors is %inherently
treated as a valid and supported usage. We then capture the oracle's exact execution status, standard outputs, and any file system state changes. This combination of the input command and the captured oracle execution result forms a complete, ground-truth behavioral test case, effectively eliminating the need for manual test verification. If a generated command fails unexpectedly during this stage, the error trace is fed back to the LLM to refine the command until a valid test template is established. 

Crucially, this fuzzing process acts as a final rigorous quality filter. If the LLM consistently fails to find a successful execution instance for any documented command pattern, the entire repository is discarded from our benchmark. Through this multi-stage filtering process,
%—from the initial 15,000 candidates to dynamic execution, manual disambiguation, and final fuzzing verification—
we ultimately yield our final dataset of 94 high-quality tasks. For the successfully verified command patterns in these remaining repositories, we synthesize exactly 50 end-to-end test cases to balance %evaluation comprehensiveness and computational efficiency.
evaluation coverage and computational cost. 
%These instances deliberately encompass both valid (positive) inputs to test normal execution paths, and invalid or edge-case (negative) inputs—such as malformed formats, out-of-bound values, or malicious strings—to rigorously evaluate whether the generated software accurately replicates the error-handling mechanisms and robustness of the original oracle.
These instances deliberately include valid positive inputs for normal execution paths and invalid or edge-case negative inputs, such as malformed formats, out-of-bounds values, or malicious strings. This design allows us to rigorously evaluate whether the generated software accurately reproduces the error handling behavior and robustness of the original oracle.

\subsubsection{Standardized task prompt construction}
The final evaluation prompt %must
is designed to simulate a realistic %, greenfield 
and new software development requirements without leaking internal implementation details. The prompt comprises three anonymized components: (1) the sanitized functional context derived from the \texttt{README}; (2) the complete, iteratively extracted ``\texttt{--help}'' documentation, serving as the strict external interface specification; and (3) one verified successful execution example for each command pattern, which is obtained from the fuzzing stage and used to clearly demonstrate the expected behavior. To %prevent any risk 
reduce the risk of data contamination during %the 
evaluation (e.g., models retrieving the original source code from the internet or their training corpora), %our 
the entire pipeline inherently embeds multi-layer defense mechanisms. 
%Specifically, the evaluation occurs in completely network-isolated Docker containers. 
Specifically, the evaluation is conducted in Docker containers without network access. During the prompt construction described above, we systematically %scrub 
remove all identifying metadata such as author names and GitHub links. Furthermore, %the 94 repositories selected in the curation phase were predominantly created 
most of the 94 repositories selected in the curation phase were created after the major training cutoffs of the evaluated LLMs.
%(obtained from the fuzzing stage) to unambiguously demonstrate the expected behavior. 

%It is worth noting that, 

\subsection{Black-box differential evaluation}
\label{sec:black-box-differential-evaluation}
To evaluate the agent-generated repository $R_{test}$ without relying on predefined structural scaffolds or internal implementations, we design a black-box Differential Evaluation Engine based on isolated Docker sandboxes. The pipeline consists of three key phases:

\subsubsection{Environment initialization}
For each task, we instantiate %two identical, language-specific Docker containers. 
two identical Docker containers for the corresponding programming language. We mount the frozen oracle repository $R_{oracle}$ and the generated repository $R_{test}$ into their respective containers. After executing standard package  installation commands, we capture the initial state of the workspaces as clean snapshots. This snapshot-and-restore mechanism guarantees strict stateless isolation, avoiding file system pollution (persistent side effects) across sequential tests. We denote these restorable base environments as $E_{oracle}$ and $E_{test}$.

\subsubsection{Differential execution}
For each test case $t$ generated in the fuzzing stage, we execute it independently in both $E_{oracle}$ and $E_{test}$. Our engine monitors the execution process and records the resulting state transition as a tuple $Exec(R, t, E) \rightarrow \langle C_{ret}, \Delta S, O_{std} \rangle$, where $C_{ret}$ is the return code, $\Delta S$ represents the raw system-level side effects (e.g., file system modifications), and $O_{std}$ is the standard output stream.

\subsubsection{Multi-tiered equivalence evaluation}
Based on the captured execution states, we evaluate functional equivalence using a strict progressive %funnel.
3-step process. A test case is considered successful only if it passes all three consecutive checks:

\textbf{1) Execution reliability ($M_{code}$)}: We strictly evaluate on valid functional paths where the oracle successfully executes without errors ($C_{ret}^{oracle} = 0$). We first verify if the agent-generated tool also completes successfully without throwing unhandled exceptions ($C_{ret}^{test} = 0$). We denote the pass rate of this metric as \textbf{Exec}.

\textbf{2) Side-effect pass ($M_{diff}$)}: Many CLI tools fundamentally operate by interacting with the file system. By comparing the workspace states before and after execution, we extract the effective file system changes $\Delta S'$ (filtering out trivial artifacts like hidden caches). A test case passes this metric, hereafter referred to as \textbf{SP}, only if it executes without errors and its effective side effects perfectly match those of the oracle ($\Delta S'_{oracle} = \Delta S'_{test}$). This prevents false positives where a tool runs successfully but performs no actual operations.

\textbf{3) Behavioral equivalence ($M_{out}$)}: Only test cases that pass both the Exec and SP checks proceed to standard output comparison ($O_{std}^{test}$ versus $O_{std}^{oracle}$). Because functionally correct CLI tools may inherently present their outputs in diverse formats or stylistic layouts, strict string matching is often insufficient. Therefore, we implement three complementary matching criteria (\textbf{EM}, \textbf{FM}, and \textbf{SM}) to ensure a comprehensive and fair evaluation across different semantic levels: 
\begin{itemize}
    \item \textbf{Exact Match (EM)} demands strict string equivalence after basic whitespace normalization.
    \item \textbf{Fuzzy Match (FM)} computes a normalized Levenshtein edit distance %,
    and records a match if the string similarity exceeds a threshold $\tau$, %to tolerate 
    thereby allowing minor formatting %divergences.
    differences.
    \item \textbf{Semantic Match (SM)} employs an %LLM-as-a-judge
    LLM as a judge (GPT-5.4) to verify whether the core informational payload is equivalent while disregarding stylistic differences. To validate this metric, a large-scale human annotation study on 1,000 output pairs yielded a Cohen's Kappa coefficient of $\kappa > 0.9$, confirming strong alignment with expert judgment. Two annotators with at least 4-year software engineering experience independently labeled whether each pair was semantically equivalent from the perspective of CLI users, ignoring superficial formatting differences but requiring preservation of core information, error messages, and observable behavior. Disagreements were resolved through discussion.
\end{itemize}

% Through the seamless integration of rigorous repository curation, schema-guided task synthesis, and black-box differential testing, our methodology culminates in a comprehensive and highly scalable benchmark. Ultimately, the pipeline yields a curated dataset of 100 high-quality CLI repositories, with 50 end-to-end test commands automatically generated for each distinct command type.

% \rdhu{[Add A Table To Show The Difficulty and Category Distribution.]}
To provide a clear overview of the constructed benchmark, Table~\ref{tab:dataset_stats} summarizes the statistical distribution of the final 94 curated tasks across three key dimensions: task difficulty, programming language, and application domain. We %specifically 
focus on Python, JavaScript (Node.js), and Go to provide a representative mix of interpreted and compiled %paradigms 
languages extensively utilized in modern CLI development. Following the taxonomy established by NL2Repo-Bench~\cite{DBLP:journals/corr/abs-2512-12730}, we stratify the task difficulty into three levels based on the oracle's Lines of Code (Easy $\le 1500$, Medium $1500-4000$, Hard $\ge 4000$). % and 
We also classify the repositories into nine distinct real-world application scenarios. As %illustrated
shown in Table~\ref{tab:dataset_stats}, the final benchmark maintains a highly diverse and balanced composition, %ensuring our evaluation comprehensively reflects 
allowing the evaluation to reflect an agent's general-purpose software generation capabilities.

\begin{table}[t]
\centering
\caption{Statistical distribution of our benchmark.}
\label{tab:dataset_stats}
\resizebox{\linewidth}{!}{
\begin{tabular}{ll cr}
\toprule
\rowcolor{gray!10}\textbf{Dimension} & \textbf{Sub-category} & \textbf{Count} & \textbf{Avg. LOC} \\
\midrule
\multirow{3}{*}{\textbf{Difficulty}}
& Easy ($\le 1500$ LOC)       & 40 & 635.90 \\
& Medium ($1500 - 4000$ LOC)  & 21 & 2,687.57 \\
& Hard ($\ge 4000$ LOC)       & 33 & 18,509.97 \\
\midrule
\multirow{3}{*}{\textbf{Language}}
& Python                      & 34 & 4,473.50 \\
& JavaScript                  & 15 & 5,770.13 \\
& Go                          & 45 & 10,090.07 \\
\midrule
\multirow{9}{*}{\textbf{Domain}}
& Web Development             & 8 & 5,550.50 \\
& Testing                     & 7 & 2,300.29 \\
& Utility Libraries           & 20 & 8,567.65 \\
& Machine Learning            & 12 & 3,334.92 \\
& Data Analysis \& Processing & 11 & 2,887.27 \\
& Database Interaction        & 6 & 3,509.17 \\
& Networking Tools            & 7 & 6,384.14 \\
& Batch File Processing       & 13 & 22,078.62 \\
& System Tools                & 10 & 3,630.00 \\
\bottomrule
\end{tabular}
}
\end{table}

\section{Experimental Setup}
\label{experiment_section}
\subsection{Selected LLMs and agent frameworks}
\label{experiment_A}
To comprehensively evaluate the state-of-the-art in autonomous software generation, we select 7 cutting-edge LLMs, encompassing both leading closed-source models and highly capable open-source models: GPT-5.4, Claude-Sonnet-4.6, DeepSeek-V3.2, Qwen-3.5-plus, GLM-5, MiniMax-M2.5, and Kimi-k2.5.

To effectively evaluate these models' capabilities as software engineers, we employ two representative agent frameworks specifically designed for repository-level tasks:

$\bullet$ OpenHands (with CodeAct)~\cite{DBLP:conf/iclr/0001LSXTZPSLSTL25}: A prominent open-source agent framework that utilizes the CodeAct paradigm, allowing the LLM to iteratively execute code, interact with a bash terminal, and observe environmental feedback.

$\bullet$ Mini-SWE-Agent~\cite{mini-swe-agent}: A streamlined adaptation of the popular SWE-agent framework, specifically optimized for iterative repository construction and terminal-based debugging.

By evaluating seven models across two frameworks, we conduct a total of 14 distinct agent configurations for each of the 94 repositories in our benchmark.

\textbf{Generation settings.} To simulate a realistic autonomous software engineering scenario, we adopt an unconstrained, zero-external-feedback generation setting. For each repository, the agent is provided with the task prompt and an empty workspace. It is allowed an unrestricted number of conversational turns to freely explore, write, and execute code within its sandbox. The generation trajectory concludes entirely based on the agent's own termination signal (e.g., executing a specific ``exit'' action), with no intermediate test feedback provided by our evaluation engine. This ``single-trajectory'' paradigm aligns seamlessly with our core objective of evaluating 0-to-1 software generation, where an agent is tasked to architect and deliver a complete, runnable tool from scratch in a single session.

\subsection{Evaluation metrics and scoring mechanism}
To systematically quantify the performance of an agent-generated repository $R_{test}$ and populate our final evaluation tables, we compute final scores for the metrics defined in Section~\ref{sec:black-box-differential-evaluation}. 

First, we compute the \textbf{Build (Global Installation Success Rate)}, measuring whether $R_{test}$ can be successfully installed using standard package managers. A repository failing this fundamental prerequisite is penalized with a score of zero across all subsequent evaluation metrics. This zero-tolerance penalty reflects the most basic requirement of the intent-driven ``Vibe Coding'' paradigm: if an agent fails to deliver a structurally valid and installable package, any downstream functional evaluation becomes meaningless. Thus, the Build metric serves as the primary indicator of an agent's foundational repository planning and environment configuration capabilities.

For the successfully installed repositories, we employ a macro-averaging aggregation strategy to calculate the scores for the aforementioned metrics (Exec, SP, EM, FM, and SM) defined in Section~\ref{sec:black-box-differential-evaluation}. This ensures that command patterns with simpler logic do not disproportionately dominate the overall evaluation. Specifically, let $C$ denote the set of all identified command patterns for a given repository $R_{test}$. For each command pattern $c \in C$, let $T_{c}$ be its corresponding suite of 50 test cases. We define the pass rate of a specific command pattern under a given metric $m \in \{\text{Exec, SP, EM, FM, SM}\}$ as:
\begin{equation}
    P_m(c) = \frac{1}{|T_c|} \sum_{t \in T_c} \mathbb{I}_m(t)
\end{equation}
where $\mathbb{I}_m(t) \in \{0, 1\}$ is the binary indicator of whether test case $t$ successfully passes the specified metric check. Note that for behavioral metrics (EM, FM, SM), $\mathbb{I}_m(t)=1$ strictly requires that the test case has already passed the prerequisite side-effect consistency gate.

The final aggregated score for the repository $R_{test}$ under metric $m$ is the unweighted average across all its command patterns:
\begin{equation}
    Score_m(R_{test}) = \frac{1}{|C|} \sum_{c \in C} P_m(c)
\end{equation}

The final values reported in our experimental results represent the average $Score_m$ across all evaluated repositories.

\subsection{Implementation details}
To prevent data contamination, we explicitly disable network access within the agent sandboxes during generation, ensuring models cannot retrieve original repository implementations. Furthermore, we remove all artificial constraints—such as maximum iteration limits or token budgets—across all models and frameworks. The generation process is entirely open-ended, allowing the agent to autonomously determine when the software construction is complete and voluntarily terminate the execution.
For all auxiliary tasks within our automated pipeline—including iterative schema extraction, LLM-directed fuzzing test generation, and the Semantic Match judge—we strictly utilize GPT-5.4. To eliminate generation randomness and ensure highly deterministic evaluation outcomes, the temperature for these auxiliary GPT-5.4 API calls is uniformly set to zero. Moreover, $\tau$ is set to 0.8 in this work.

\subsection{Research questions}
We investigate the following four research questions (RQs):

\begin{itemize}[leftmargin=*]
    \item \textbf{RQ1 (Overall capabilities):} How do state-of-the-art LLMs perform in 0-to-1 CLI tool generation, and how do different agent frameworks influence their functional correctness?
    \item \textbf{RQ2 (Complexity impact):} How does the autonomous generation capability of these models scale or degrade across varying degrees of project complexity?
    \item \textbf{RQ3 (Generation efficiency):} What are the computational overheads and token consumption patterns of these agents under an unconstrained, open-ended generation setting?
    \item \textbf{RQ4 (Structural autonomy):} Given complete structural freedom, how do the repository-level designs of agent-generated software differ from human-written oracles?
\end{itemize}

\section{Experimental Result}
\label{sec:result}
In this section, we present the evaluation results to answer our four research questions. 

Before diving into the detailed RQs, it is important to contextualize the scale and statistical reliability of our findings. Under the realistic single-trajectory setting (Section~\ref{experiment_A}), each of the seven evaluated LLMs independently generated the 94 repositories across two frameworks. Because our benchmark leverages LLM-directed fuzzing to synthesize exactly 50 end-to-end test cases per command pattern, a single generated repository yields hundreds of independent test instances. Consequently, the macro-averaged scores presented in the subsequent sections are derived from a massive distribution of tens of thousands of test executions per model. 

To validate that the performance differences between models observed on this vast dataset are reliable and not just random chance, we employed McNemar's test on the pass/fail outcomes, alongside Bootstrap resampling (10,000 iterations) to estimate the 95\% Confidence Intervals (CI). Our baseline analyses confirm that the ranking of the top-tier models is statistically significant ($p < 0.05$). This proves that the differences in their aggregated scores reflect true variations in their coding capabilities, rather than sampling noise from the single generation runs, thereby establishing a rigorous foundation for the following research questions.

% ==========================================
% RQ1: Overall Performance
% ==========================================
\subsection{RQ1: Overall performance and framework impact}
% [Table 1: 主实验结果大表]
\begin{table*}[t]
\centering
\caption{Overall performance of the evaluated LLMs across two agent frameworks. The rightmost columns and bottom row present macro-averages. Best results are highlighted in \textbf{bold}.}
\label{tab:main_results}
\setlength{\tabcolsep}{2pt}
\resizebox{\textwidth}{!}{
\begin{tabular}{l cccccc c cccccc c cccccc}
\toprule
\multirow{2}{*}{\textbf{Model}} & \multicolumn{6}{c}{\textbf{OpenHands}} & & \multicolumn{6}{c}{\textbf{Mini-SWE-Agent}} & & \multicolumn{6}{c}{\textbf{Average (Both Frameworks)}} \\
\cmidrule{2-7} \cmidrule{9-14} \cmidrule{16-21}
& \textbf{Build} & \textbf{Exec} & \textbf{SP} & \textbf{EM} & \textbf{FM} & \textbf{SM} & & \textbf{Build} & \textbf{Exec} & \textbf{SP} & \textbf{EM} & \textbf{FM} & \textbf{SM} & & \textbf{Build} & \textbf{Exec} & \textbf{SP} & \textbf{EM} & \textbf{FM} & \textbf{SM} \\
\midrule
GPT-5.4             & 79.79 & 59.66 & 38.83 & 21.47 & 38.55 & 28.88 & & 85.11 & 61.05 & 51.90 & 23.85 & 40.74 & 31.82 & & 82.45 & 60.35 & 45.36 & 22.66 & 39.65 & 30.35 \\
Claude-Sonnet-4.6   & 37.23 & 18.42 & 12.34 & 5.95 & 9.36 & 7.21 & & 52.13 & 31.67 & 28.15 & 11.77 & 21.03 & 14.02 & & 44.68 & 25.04 & 20.24 & 8.86 & 15.20 & 10.61 \\
DeepSeek-V3.2       & 80.85 & 61.82 & 41.53 & 21.48 & 37.72 & 27.44 & & 72.34 & 60.08 & 50.67 & 21.25 & 40.33 & 27.41 & & 76.60 & 60.95 & 46.10 & 21.37 & 39.02 & 27.43 \\
Qwen-3.5-plus       & 73.40 & 58.96 & 42.71 & 23.50 & 38.34 & 28.86 & & 76.60 & 61.34 & 49.38 & 24.15 & 39.99 & 30.75 & & 75.00 & 60.15 & 46.05 & 23.83 & 39.16 & 29.81 \\
GLM-5               & 76.60 & 59.03 & 40.05 & 22.07 & 37.43 & 28.54 & & 78.72 & 63.47 & 53.44 & 26.69 & 43.97 & 33.23 & & 77.66 & 61.25 & 46.75 & 24.38 & 40.70 & 30.89 \\
MiniMax-M2.5        & 79.79 & 66.10 & 42.01 & 22.98 & 42.27 & 30.11 & & \textbf{95.74} & \textbf{80.12} & 68.34 & 31.22 & 51.96 & 39.07 & & 87.77 & 73.11 & 55.18 & 27.10 & 47.12 & 34.59 \\
Kimi-k2.5           & \textbf{90.43} & \textbf{69.65} & \textbf{50.53} & \textbf{29.68} & \textbf{46.83} & \textbf{36.41} & & \textbf{95.74} & 79.17 & \textbf{69.37} & \textbf{43.00} & \textbf{57.47} & \textbf{51.16} & & \textbf{93.09} & \textbf{74.41} & \textbf{59.95} & \textbf{36.34} & \textbf{52.15} & \textbf{43.78} \\
\midrule
\rowcolor{gray!10}
\textit{Average}    & \textit{74.01} & \textit{56.23} & \textit{38.29} & \textit{21.02} & \textit{35.79} & \textit{26.78} & & \textit{79.48} & \textit{62.41} & \textit{53.04} & \textit{25.99} & \textit{42.21} & \textit{32.50} & & \textit{76.75} & \textit{59.32} & \textit{45.66} & \textit{23.50} & \textit{39.00} & \textit{29.64} \\
\bottomrule
\end{tabular}
}
\end{table*}

\begin{figure}[t]
    \centering
    \includegraphics[width=\linewidth]{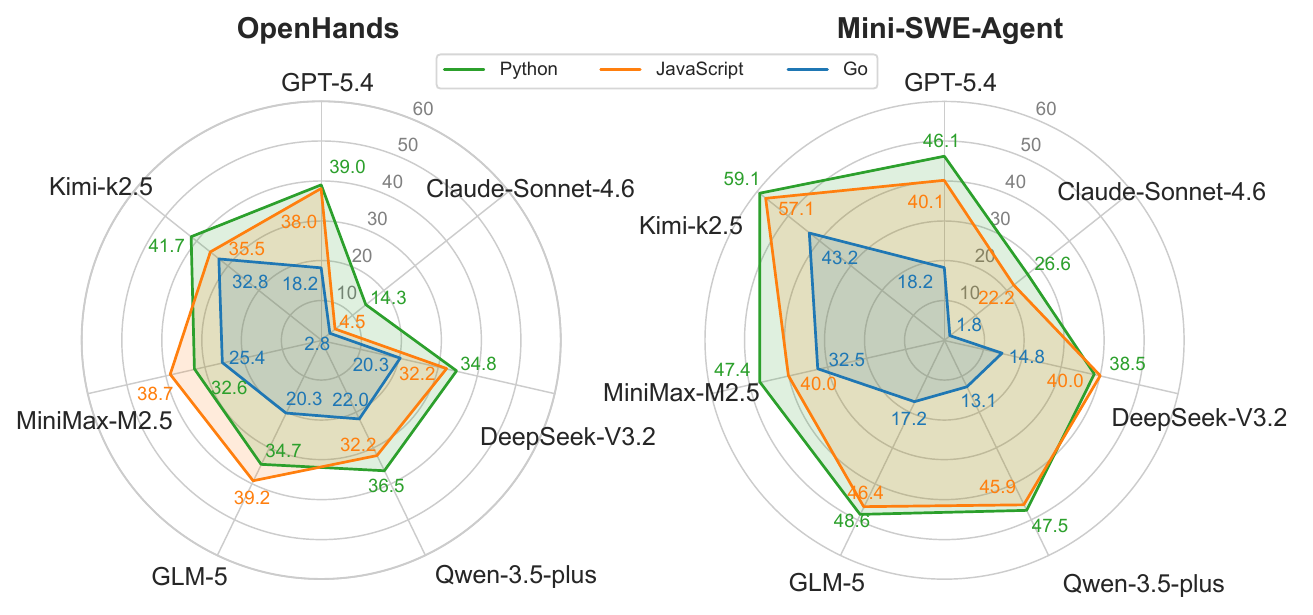}
    \caption{Performance comparison of seven models across three programming languages. The radar charts illustrate the varying capabilities of each model, with the radial axes representing Semantic Match ($SM$) scores.}
    \label{fig:language}
\end{figure}

To answer RQ1, we evaluate seven state-of-the-art LLMs across two distinct agent frameworks. As detailed in Table~\ref{tab:main_results}, the results reveal a clear performance hierarchy. Kimi-k2.5 emerges as the frontrunner, achieving the highest average Semantic Match (SM) score of 43.78\%, followed by MiniMax-M2.5 (34.59\% SM). A middle tier consists of GPT-5.4, Qwen-3.5-plus, and GLM-5, performing around 29\% to 31\% SM. Regarding framework impact, we observe a consistent trend: models generally achieve higher scores using Mini-SWE-Agent compared to OpenHands. This suggests that the action space and interface design of the agent framework substantially affect task completion in 0-to-1 generation scenarios.

Conversely, Claude-Sonnet-4.6 exhibits notably lower performance (averaging 10.61\% SM), primarily bottlenecked at the initial Build stage. We note that our benchmark strictly provides no post-submission test feedback. In this setting, we observe a distinct behavioral pattern in Claude-Sonnet-4.6: it frequently concludes the generation trajectory immediately after writing the initial code. For instance, across multiple Go projects, the agent submits the repository without invoking standard compilation checks like \texttt{go build} or \texttt{go mod init}. Consequently, these repositories fail the initial build checks in our evaluation. In contrast, log analysis reveals that models like Kimi-k2.5 typically engage in more extensive in-container Bash validations before issuing the termination signal. This suggests that the performance gap at the Build stage is partially driven by differences in the models' tendency to autonomously verify their code prior to submission.

Beyond individual model capabilities, our progressive evaluation funnel exposes fundamental bottlenecks in current autonomous workflows. Across all models, there is a steep degradation from the Build stage (average 76.75\%) to Execution Reliability (59.32\%), and a further drop at the Side-Effect Pass (45.66\%) gate. Ultimately, performance plunges when evaluated by Exact Match (EM, 23.50\%). This massive drop demonstrates that while agents successfully write syntactically correct and executable code in many instances, they struggle to perfectly replicate the exact string outputs of the human oracle. However, the notable recovery observed in the FM score (39.00\%) and SM score (29.64\%) validates our methodological design: agents frequently generate functionally correct and semantically equivalent CLI tools that are unfairly penalized by stringent EM metrics.

A deeper analysis of the programming language distribution (Figure~\ref{fig:language}) further reveals a pronounced language bias inherent in current LLMs. As depicted in the radar charts, almost all evaluated models exhibit larger coverage areas for interpreted languages like Python and JavaScript, while Golang emerges as a severe bottleneck (represented by the constricted inner green polygons). It suggests that models frequently fail to navigate Golang's strict type matching and rigid compilation constraints, often generating code that fails to build. 

Furthermore, the choice of agent framework appears to influence generation outcomes. We observe that the streamlined, bash-centric Mini-SWE-Agent consistently achieves higher scores than the more complex OpenHands framework. This suggests that minimalist interfaces may naturally align better with terminal-centric workflows. A closer inspection of individual agent trajectories supports this observation. For instance, when attempting to resolve a missing library reference in the \texttt{mol-lang} repository, OpenHands agents frequently invoke browser tools, only to be flooded with multi-page HTML that overwhelms their context windows. Conversely, Mini-SWE-Agent directly reads the brief terminal \texttt{stderr}, executes the package manager, and successfully proceeds.

\begin{tcolorbox}[myfindingbox]
\textbf{Finding 1:} Despite the promising capabilities of top-tier models, their steep performance degradation across the evaluation funnel and severe struggles with compiled languages cap the highest overall success rate at 43.78\%. This highlights that end-to-end, 0-to-1 software generation remains a highly challenging frontier.
\end{tcolorbox}

% ==========================================
% RQ2: Task Complexity
% ==========================================
\subsection{RQ2: Impact of task complexity}
% [Figure 1: 难度折线图/柱状图占位]
\begin{figure}[t]
    \centering
    \includegraphics[width=\linewidth]{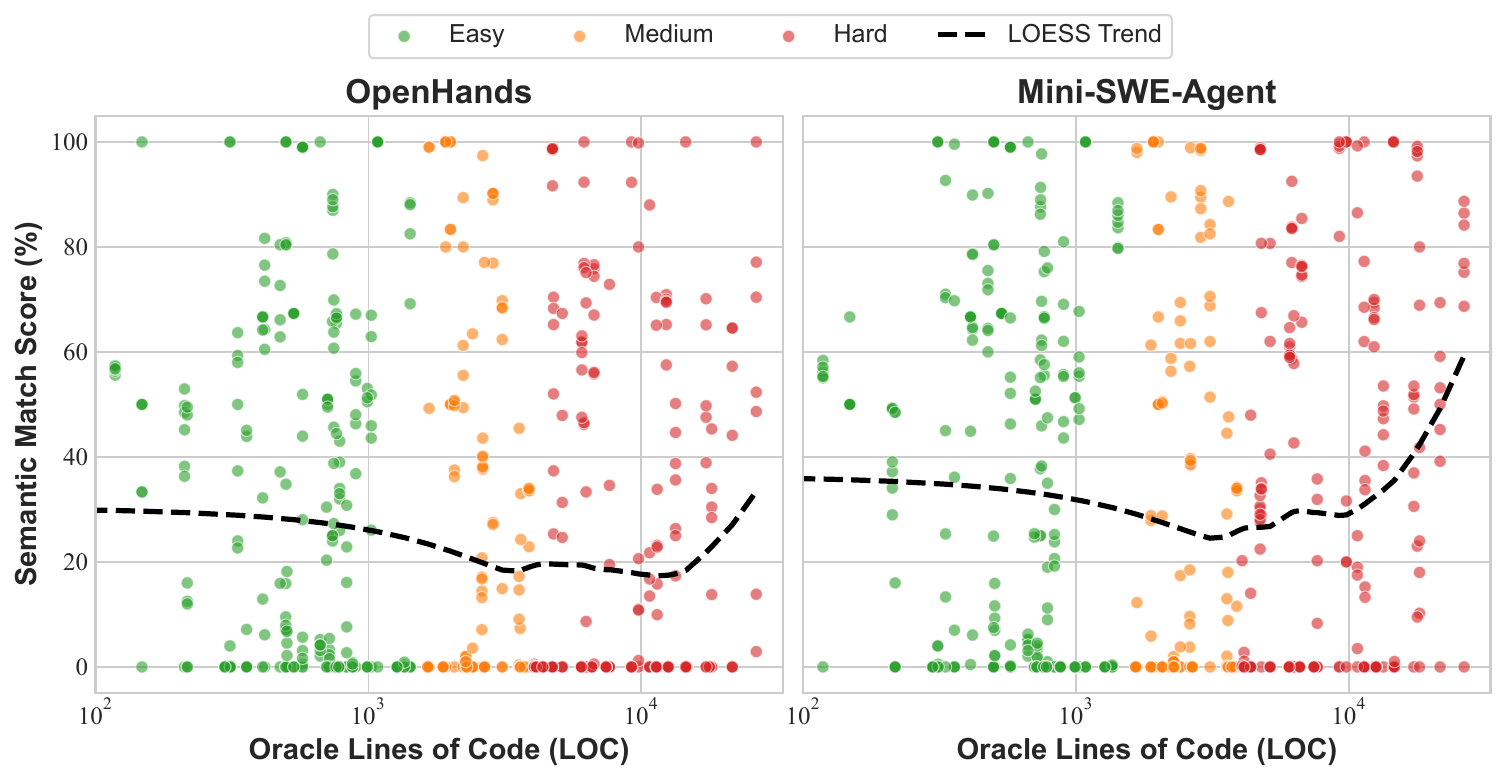}
% \vspace{-2em}
    \caption{Correlation between repository complexity and agent performance. Each scatter point represents an individual test case. The color of the points denotes the difficulty level: Green for Easy, Orange for Medium, and Red for Hard. The dashed black line represents the LOESS trend line.}
    \label{fig:complexity}
% \vspace{-1em}
    
\end{figure}

To answer RQ2, we investigate how the software generation capability of these agents scales across varying degrees of project complexity. Figure~\ref{fig:complexity} illustrates the relationship between the oracle repository's size (measured in Lines of Code, LOC) and the corresponding Semantic Match (SM) scores. Intuitively, one might expect a strict negative correlation: as the repository grows larger, the agent's performance should monotonically degrade due to expanding context windows and complex dependency resolutions. However, the LOESS trend lines reveal a non-monotonic (U-shaped) trajectory.

In the transition from Easy ($<$1500 LOC) to Medium (1500$-$4000 LOC) repositories, we observe a steep decline in performance. The overall average SM score drops from 32.66\% to 25.08\%, with competitive models like GPT-5.4 plummeting from 34.60\% to 21.90\%. To understand this degradation, we analyze the structural characteristics of these repositories. Our data reveal that while Easy tasks are predominantly flat, single-file scripts (averaging 1.62 files), Medium tasks demand multi-file, cross-directory modularization (averaging 2.33 files and 1.14 subdirectories). This structural leap appears to force the agent to handle complex relative imports and cross-file state management. The sharp drop in scores suggests that current LLMs struggle significantly when transitioning from localized coding to repository-level architectural planning.

Surprisingly, as the LOC scales into the Hard category ($>$4000 LOC), the average SM score rebounds to 28.87\%. To unpack this counterintuitive phenomenon, we investigate the contextual richness associated with massive repositories. A quantitative analysis of our dataset reveals that the scale of a project strongly correlates with the comprehensiveness of its documentation: the average length of the \texttt{README} in Hard tasks (11,728 characters) is nearly three times that of Easy tasks (4,501 characters). In our intent-driven, 0-to-1 generation setting, this meticulously structured documentation provides a highly unambiguous specification. We hypothesize that this ``specification dividend'' significantly offsets the raw implementation difficulty. While agents likely still struggle to perfectly architect the underlying logic of massive systems, the exceptionally rich and explicit documentation may enable them to more accurately extract and mirror the expected CLI command schemas, thereby achieving a higher SM score.

\begin{tcolorbox}[myfindingbox]
    \textbf{Finding 2:} Agent performance exhibits a non-monotonic, U-shaped trend with repository complexity. The initial sharp decline highlights models' severe struggles with multi-file architectural planning. However, the performance rebound in massive repositories suggests that highly comprehensive, unambiguous documentation significantly counteracts raw implementation difficulty, allowing agents to accurately reconstruct semantic schemas.
\end{tcolorbox}
% ==========================================
% RQ3: Cost & Efficiency
% ==========================================
\subsection{RQ3: Agent cost and generation Efficiency}
% [Table 2: 开销统计表]
\begin{table*}[t]
\centering
\caption{Generation overhead and estimated API cost per repository, categorized by LLM and agent framework.}
% \vspace{-1em}
\label{tab:overhead}
\resizebox{\textwidth}{!}{
\begin{tabular}{ll ccc c rrr c rrr c rrr}
\toprule
\multirow{2}{*}{\textbf{Model}} & \multirow{2}{*}{\textbf{Framework}} & \multicolumn{3}{c}{\textbf{Steps}} & & \multicolumn{3}{c}{\textbf{Prompt Tokens (K)}} & & \multicolumn{3}{c}{\textbf{Comp. Tokens (K)}} & & \multicolumn{3}{c}{\textbf{Cost (\$)}} \\
\cmidrule{3-5} \cmidrule{7-9} \cmidrule{11-13} \cmidrule{15-17}
& & \textbf{Avg} & \textbf{Min} & \textbf{Max} & & \textbf{Avg} & \textbf{Min} & \textbf{Max} & & \textbf{Avg} & \textbf{Min} & \textbf{Max} & & \textbf{Avg} & \textbf{Min} & \textbf{Max} \\
\midrule
\multirow{2}{*}{GPT-5.4}
& OpenHands & 12.03 & 5 & 26 & & 159.43 & 43.75 & 722.91 & & 9.07 & 1.69 & 25.89 & & 0.53 & 0.15 & 1.94 \\
& Mini-SWE-Agent & 2.33 & 2 & 6 & & 17.68 & 2.81 & 82.45 & & 3.21 & 0.30 & 8.26 & & 0.09 & 0.01 & 0.26 \\
\midrule
\multirow{2}{*}{Claude-Sonnet-4.6}
& OpenHands & 7.48 & 1 & 34 & & 196.74 & 31.84 & 1,917.41 & & 7.09 & 0.12 & 81.78 & & 0.70 & 0.10 & 6.15 \\
& Mini-SWE-Agent & 7.17 & 2 & 26 & & 92.01 & 8.96 & 483.82 & & 7.58 & 0.51 & 39.36 & & 0.39 & 0.03 & 1.77 \\
\midrule
\multirow{2}{*}{DeepSeek-V3.2}
& OpenHands & 60.87 & 24 & 132 & & 2,031.91 & 352.04 & 9,851.36 & & 18.14 & 4.30 & 40.04 & & 0.58 & 0.10 & 2.77 \\
& Mini-SWE-Agent & 39.64 & 14 & 112 & & 802.97 & 43.56 & 5,626.74 & & 15.66 & 1.82 & 51.70 & & 0.23 & 0.01 & 1.60 \\
\midrule
\multirow{2}{*}{Qwen-3.5-plus}
& OpenHands & 34.03 & 2 & 130 & & 1,099.83 & 1.23 & 5,599.58 & & 14.20 & 2.08 & 47.24 & & 0.47 & 0.03 & 2.35 \\
& Mini-SWE-Agent & 35.40 & 8 & 131 & & 911.92 & 18.97 & 6,907.64 & & 18.90 & 1.32 & 71.86 & & 0.41 & 0.01 & 2.88 \\
\midrule
\multirow{2}{*}{GLM-5}
& OpenHands & 25.67 & 1 & 119 & & 229.16 & 0.23 & 1,073.98 & & 9.62 & 0.23 & 37.56 & & 0.26 & 0.00 & 1.19 \\
& Mini-SWE-Agent & 39.19 & 2 & 116 & & 221.77 & 9.45 & 1,005.59 & & 13.48 & 0.98 & 57.60 & & 0.26 & 0.01 & 1.19 \\
\midrule
\multirow{2}{*}{MiniMax-M2.5}
& OpenHands & 88.02 & 20 & 276 & & 4,206.61 & 285.08 & 23,257.85 & & 29.77 & 3.82 & 93.59 & & 1.30 & 0.09 & 7.08 \\
& Mini-SWE-Agent & 68.48 & 10 & 153 & & 2,412.56 & 21.37 & 9,503.50 & & 33.32 & 1.04 & 105.40 & & 0.76 & 0.01 & 2.93 \\
\midrule
\multirow{2}{*}{Kimi-k2.5}
& OpenHands & 47.38 & 4 & 130 & & 1,317.52 & 24.92 & 6,844.40 & & 23.40 & 1.54 & 238.72 & & 0.86 & 0.03 & 4.33 \\
& Mini-SWE-Agent & 49.38 & 3 & 193 & & 337.53 & 6.04 & 4,883.85 & & 38.28 & 0.46 & 342.21 & & 0.32 & 0.00 & 3.96 \\
\bottomrule
\end{tabular}
}
% \vspace{-1em}
\end{table*}

% [Figure 2: 散点图占位 - 成本 vs 成功率]
\begin{figure}[t]
    \centering
    \includegraphics[width=\linewidth]{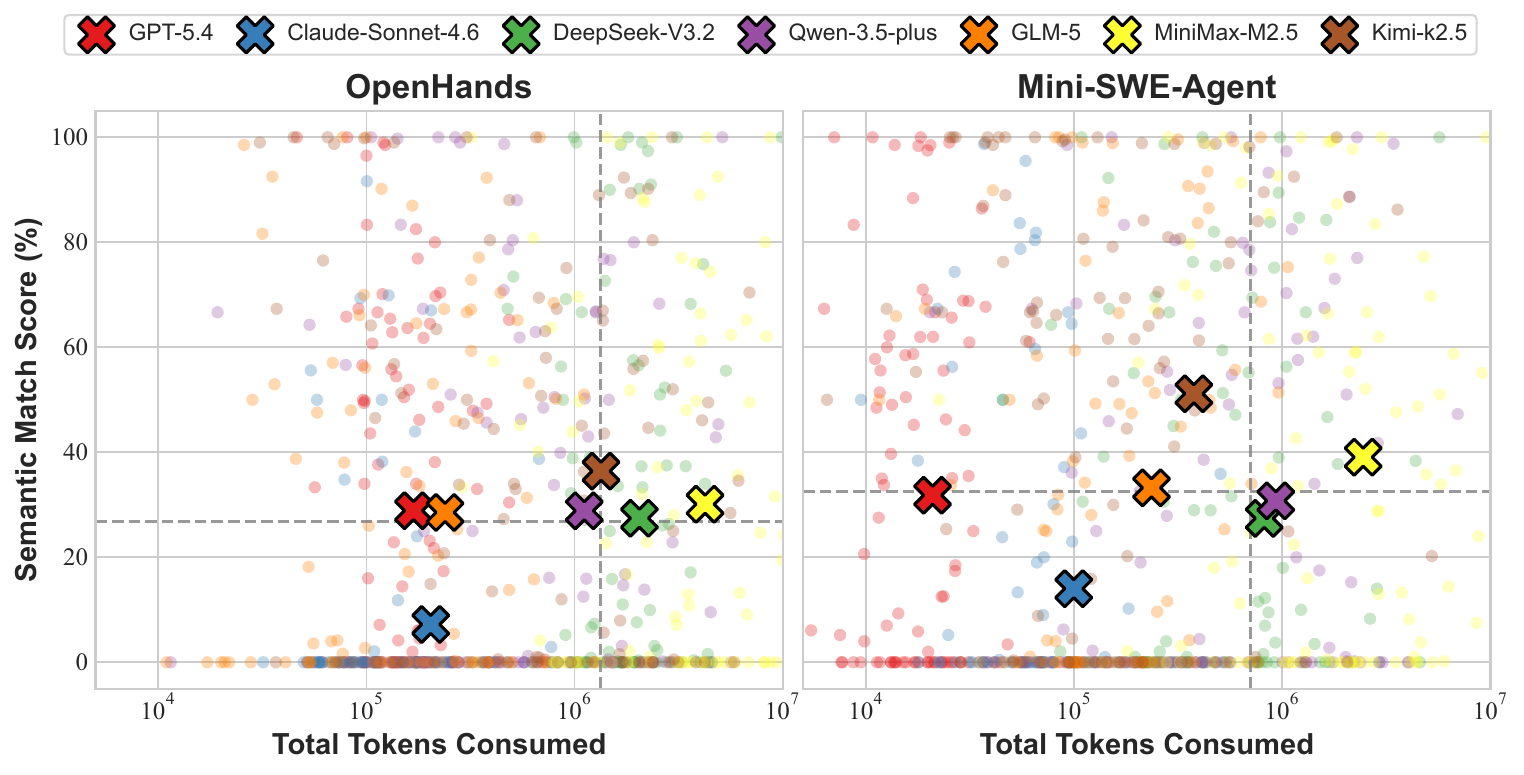}
% \vspace{-2em}
    \caption{Cost-effectiveness analysis of LLM agents across different frameworks. The scatter plots illustrate the trade-off between task performance and execution cost. Small translucent dots represent individual task runs, while the large cross markers denote the centroid for each model. The dashed gray lines indicate the overall mean score and mean token consumption, dividing the space into four quadrants.}
    \label{fig:cost_tradeoff}
    % (Semantic Match Score, y-axis)
    % (Total Tokens Consumed, x-axis)
    % (average performance and cost)
% \vspace{-1em}
    
\end{figure}
% Because we removed artificial iteration limits, the agents exhibit significant variance in how they utilize computational resources. Table \ref{tab:overhead} and Figure \ref{fig:cost_tradeoff} detail these differences.
% \textbf{Resource Utilization vs. Success:} 
% %[TODO: 结合散点图分析，谁是“性价比之王”（花最少的 Token 办成最多的事）？谁是“大力出奇迹”（狂耗 Token 但成功率高）？]
% \textbf{The "Infinite Loop" Phenomenon:} 
% %[TODO: 分析是否有模型因为没有轮次限制而陷入了死循环（比如 Avg Steps 特别高但得分很低），这说明该模型缺乏自我反思和主动终止（Autonomous Termination）的能力。]
To answer RQ3, we analyze the cost-effectiveness of different LLMs by plotting their Semantic Match Scores against their total token consumption, as shown in Figure~\ref{fig:cost_tradeoff}. The dashed lines represent the average score and token usage, dividing the performance space into four quadrants. Ideally, a highly capable agent should fall into the top-left quadrant, achieving above-average scores with below-average token costs. 

Across both frameworks, GPT-5.4 and GLM-5 consistently demonstrate this optimal behavior. GPT-5.4, in particular, emerges as a highly efficient engine. In the Mini-SWE-Agent framework, it achieves an average SM score of 31.82\% while requiring only 2.33 interaction steps and consuming less than 21K tokens per task (costing a mere \$0.09). This suggests that GPT-5.4 possesses strong zero-shot reasoning capabilities, generating accurate CLI schemas without relying on extensive, token-heavy trial-and-error loops. Similarly, Kimi-k2.5 stands out as the overall best performer. It achieves the highest SM score (51.16\%) with a moderate token footprint (roughly 375.8K tokens and 49.38 steps) in the Mini-SWE-Agent framework, effectively dominating the top-left quadrant.

Conversely, the right half of the scatter plots reveals a phenomenon of ``diminishing returns'' in agent trajectories. Models like Minimax-M2.5 and DeepSeek-V3.2 frequently fall into the right-side quadrants. For example, within OpenHands, Minimax-M2.5 consumes a staggering average of 4.21 million tokens and engages in 88.02 steps per task, yet fails to achieve top-tier performance (scoring only 30.11\% SM). This extreme token consumption and interaction count typically indicate ``thrashing''—situations where the agent becomes trapped in repetitive debugging cycles or generates overly verbose, unhelpful commands without making actual progress toward task resolution. On the other extreme, Claude-Sonnet-4.6 consistently occupies the bottom-left quadrant. Its remarkably low token usage (averaging roughly 92K tokens in Mini-SWE-Agent and 197K in OpenHands) and minimal interaction count (averaging 7.58 and 7.09 steps, respectively), coupled with the lowest semantic scores, imply a distinct ``fail fast'' tendency. As previously observed, this behavior likely stems from Claude's lack of extensive self-verification and debugging loops; it tends to produce an initial solution and terminate the trajectory prematurely, rather than engaging in the iterative troubleshooting necessary for 0-to-1 software construction.

\begin{tcolorbox}[myfindingbox]
    \textbf{Finding 3:} Higher token consumption does not equate to better task resolution. While certain models experience diminishing returns due to costly debugging loops, efficient models like GPT-5.4 achieve highly competitive performance with minimal tokens and interaction steps, demonstrating superior cost-effectiveness.
\end{tcolorbox}

% ==========================================
% RQ4: structural Autonomy
% ==========================================
\subsection{RQ4: Structural Autonomy and Diversity}
% [Table 3: 架构统计表]
% \input{tables/rq4}
% [Figure 3: 箱线图占位 - 文件数分布]
\begin{figure}[t]
    \centering
    \includegraphics[width=0.48\textwidth]{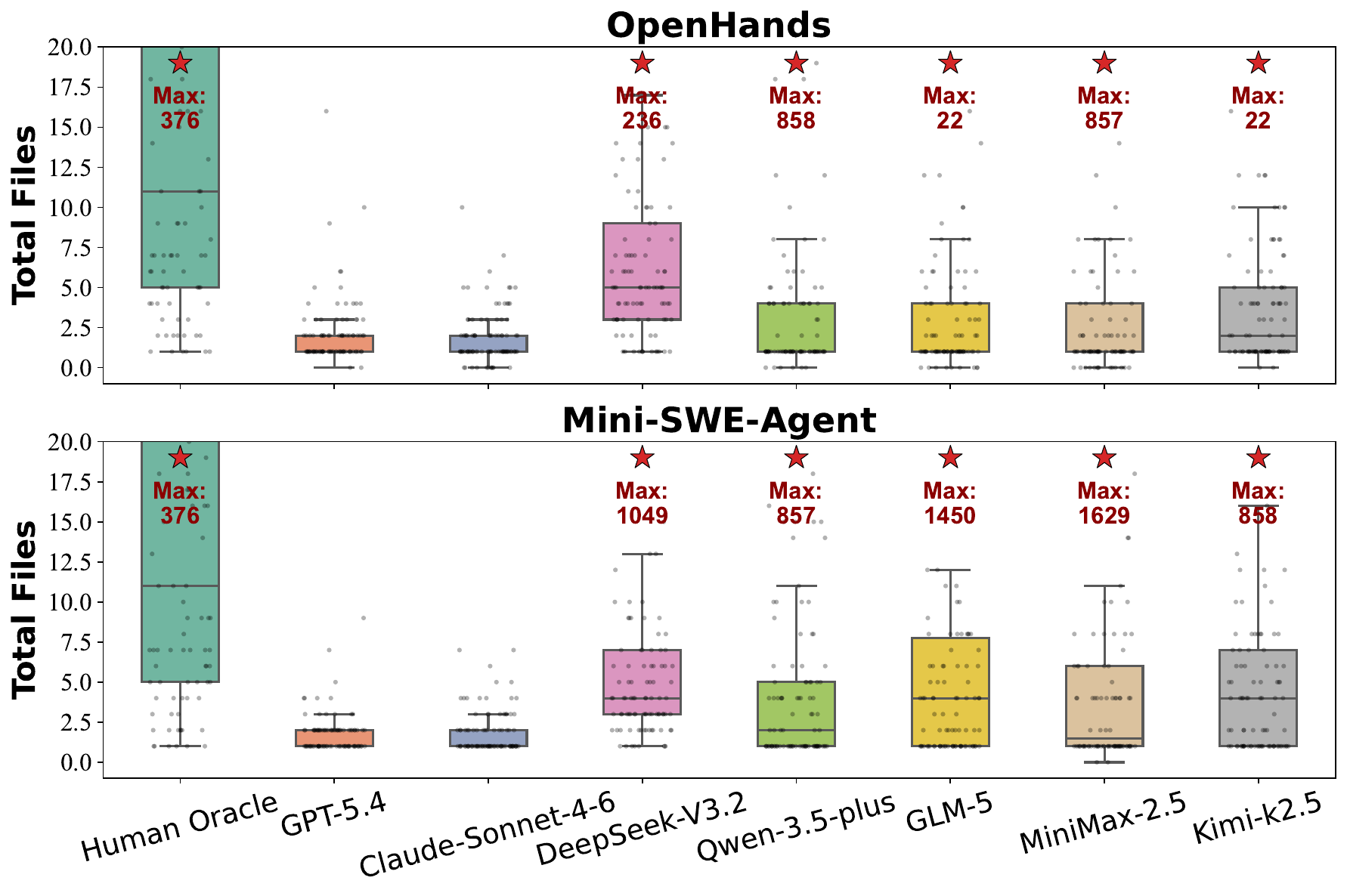}
% \vspace{-2em}
    \caption{Distribution of total files generated by the human Oracle and various LLMs. The box plots, combined with overlaid strip plots, show the median and quartile distributions while highlighting individual task runs as semi-transparent dots. The y-axis is truncated at 20 files. Extreme outliers are annotated with red stars and their maximum values at the top.}
    \label{fig:rq4}
% \vspace{-1.5em}
\end{figure}

To answer RQ4, we investigate how different LLMs organize repository structures and manage workspace complexity when granted full structural autonomy. Figure~\ref{fig:rq4} compares the file count distributions of agent-generated repositories against the human Oracle.

\textbf{Human Modularity vs. Agent Monolithic Preference.} 
The evident trend is the divergence in structural design between human developers and autonomous agents. The human Oracle exhibits a broad distribution, reflecting standard software engineering practices where code is modularized into distinct components. In contrast, all evaluated LLMs demonstrate a strong preference for monolithic structures, with their medians tightly clustered between 1 and 3 files. This indicates a shared behavioral strategy among LLMs: centralizing logic into a single or very few files. From an agentic perspective, this monolithic approach seems to be a practical adaptation to minimize cross-file dependency issues (e.g., \texttt{ImportError}) and to keep the entire system state easily accessible within the model's limited context window.

\textbf{Workspace Management and Debugging Behaviors.} 
Beyond the median values, the maximum file counts (annotated with red stars) reveal diverse workspace management behaviors during the generation process. Models like GPT-5.4 and Claude-Sonnet-4.6 maintain strictly low file counts across all tasks (maximums of 16 and 10, respectively). Conversely, several other models (e.g., DeepSeek-V3.2, Qwen-3.5-plus, and MiniMax-M2.5) occasionally exhibit extreme file sprawl, generating hundreds or even thousands of files (e.g., up to 1,629 files). Upon inspecting these repositories, we found that this extreme sprawl does not represent manually written source code, but rather the inclusion of massive local dependency directories (e.g., \texttt{node\_modules} or local virtual environments) directly within the workspace. This indicates that while some models carefully manage dependencies globally or outside the target directory, others resort to heavy local installations during their execution. Furthermore, this behavior is heavily influenced by the environment. For instance, GLM-5 and Kimi-k2.5 maintain compact workspaces under OpenHands but exhibit severe dependency sprawl under Mini-SWE-Agent, suggesting that the underlying framework's prompt structure and feedback mechanisms impact agents' file system management. Interestingly, we also observe that a notable proportion of cases generate exactly zero valid files\footnote{These ``zero-file'' cases primarily result from unauthorized workspace relocation (agents writing to hallucinated absolute paths) or premature termination due to safety filters. Detailed case analyses are available in our repository.}, particularly within the high-freedom OpenHands framework.

% Unlike traditional white-box evaluations that force agents to complete pre-defined files, our black-box setting allows agents to design the software structure from scratch. Table \ref{tab:structure} and Figure \ref{fig:arch_boxplot} compare these generated structures against the human Oracles.
% \textbf{Monolithic vs. Modular Design:} 
%[TODO: 分析大模型倾向于写“单体大文件”（比如把所有代码塞进一个 main.py，导致 Avg Files 很少），还是像人类一样进行模块化拆分（Modularization）？]
% \textbf{Directory Depth:} 
%[TODO: 分析大模型是否会建立复杂的文件夹层级（如 src/, utils/, tests/），还是倾向于扁平化结构（Flat structure）？这能深刻反映大模型的“系统级工程思维”。]

\begin{tcolorbox}[myfindingbox]
    \textbf{Finding 4:} While human developers naturally adopt modular structures, LLMs prefer monolithic designs (typically 1-3 files) to simplify context management. Furthermore, models exhibit vastly different workspace management strategies, ranging from strict in-place editing to massive local dependency sprawl, a behavior that is also highly sensitive to the chosen framework.
\end{tcolorbox}

\section{Discussion}
% \section{Discussion}
\label{sec:discussion}

% While our quantitative results in Section~\ref{sec:result} highlight the functional success rates of various LLM agents, true autonomous software engineering requires looking beyond mere correctness. In this section, we unpack typical failure modes through quantitative and qualitative case studies, and discuss the broader implications of our findings.

\begin{figure*}[t]
    \centering
    \includegraphics[width=0.9\textwidth]{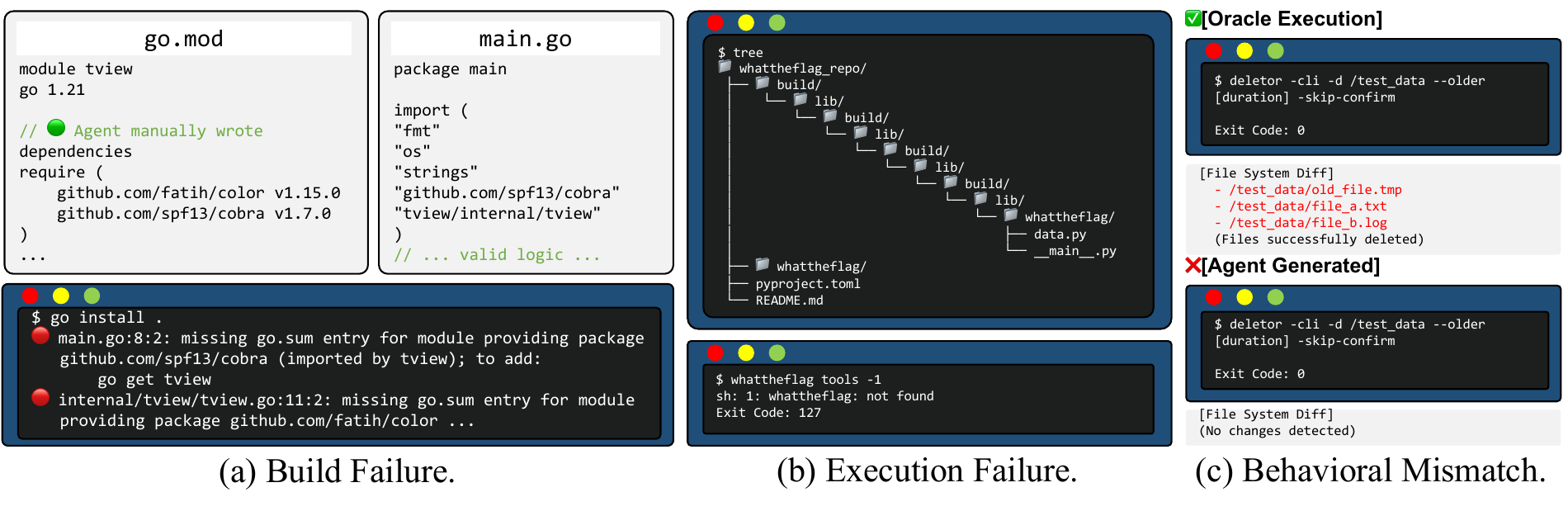}
    \caption{Representative failure cases of LLM agents in CLI tool development.}
    \label{fig:case_study}
\end{figure*}

\subsection{Case Study: Phase-based Failure Analysis}
% While our quantitative results in Section~\ref{sec:result} highlight the functional success rates of various LLM agents, true autonomous software engineering requires looking beyond mere correctness.
% We display three typical failure cases through quantitative and qualitative case studies.

\textbf{Phase 1: Build failures (23.3\% of trajectories).} Nearly a quarter of all runs fail to produce a structurally installable package. Figure~\ref{fig:case_study} (a) demonstrates a prevalent ``build failure'' driven by a \textit{text-to-file bias}. In this Go project scenario, the agent generated the correct logic but attempted to manually hardcode dependencies into \texttt{go.mod} instead of executing standard commands like \texttt{go mod tidy}. Lacking ``environmental intuition,'' agents default to static text manipulation rather than interacting with dynamic toolchains.

\textbf{Phase 2: Execution failures (13.2\% of trajectories).} Among repositories that install successfully, 13.2\% crash entirely during execution. Figure~\ref{fig:case_study} (b) illustrates a catastrophic workspace mismanagement scenario. While attempting to build a Python tool, the agent trapped itself in a recursive directory generation loop, creating deeply nested \texttt{build/lib/} structures. This chaotic file system state broke the package's entry points, highlighting a critical spatial blindness when diagnosing corrupted environments.

\textbf{Phase 3: Behavioral mismatch (6.7\% of trajectories).} Perhaps the most deceptive failure mode occurs when a tool executes flawlessly (Exit Code 0) but produces no system-level side effects (SM=0). As shown in Figure~\ref{fig:case_study} (c), the agent-generated \texttt{deletor} tool ran without crashing but failed to remove the target files. This silent failure firmly validates the necessity of our multi-dimensional evaluation approach, proving that ``running without crashing'' does not equate to task completion.

\subsection{Implications of findings}

% In this section, we discuss the implications of our work for researchers and developers.

\noindent \textbf{Researchers:} Our research demonstrates that evaluating autonomous agents requires moving beyond static code analysis. Specifically:
\begin{itemize}[leftmargin=*]
    \item \textbf{Evaluation paradigms.} The significant gap between exact match, semantic match, and execution success proves that traditional string-based evaluations are insufficient for agentic tasks. End-to-end behavioral validation must become the new standard. Therefore, it is imperative to propose novel equivalence metrics and methods that can accurately capture actual software behaviors.
    \item \textbf{Agent architecture.} The prevalent issues of ``thrashing'' and catastrophic workspace mismanagement highlight a structural gap. Future research must equip agents with better spatial awareness of the file system and self-reflection mechanisms to break out of unproductive debugging loops.
\end{itemize}

\noindent \textbf{Developers:} Agentic software generation is a new paradigm that introduces unique challenges in code structure and quality. Based on our findings, we conclude the following insights and takeaways for developers using LLM agents:
\begin{itemize}[leftmargin=*]
    \item \textbf{Structural scaffolding.} Since agents universally prefer monolithic designs and struggle with complex dependency resolutions, developers should provide explicit structural scaffolding or mandate the use of standardized CLI frameworks (e.g., \texttt{Click} or \texttt{Cobra}) to guide generation.
    \item \textbf{Defensive programming.} While agent-generated tools often execute faster than human baselines due to their stripped-down, goal-oriented coding style, they severely lack defensive programming practices. Developers must treat these outputs as functional prototypes, actively reinforcing them with robust error handling.
    \item \textbf{Tailored ``vibe coding''.} Developers must adopt tailored interaction strategies: mandate explicit self-testing steps for rapid-delivery models (e.g., Claude), while proactively monitoring verification-driven agents (e.g., Kimi) to prevent infinite loops.
\end{itemize}

\subsection{Threats to validity}
% We identify four main threats to the validity of our study:
\begin{itemize}[leftmargin=*]
 \item \textbf{Dataset Scale and Selection.} Our dataset consists of 94 curated task instances. This scale represents a deliberate trade-off: we prioritized a rigorously cleansed and manually inspected dataset over a larger, automated collection to ensure high-quality and unambiguous task specifications.

% 2) \textbf{Evaluation Metrics.} Evaluating open-ended terminal outputs relies on proxy metrics. While we strictly enforce execution and side-effect checks, our Behavioral Equivalence metrics (Fuzzy Match and Semantic Match) serve to accommodate the diverse stylistic outputs of LLMs. Although LLM-as-a-judge approaches may introduce potential bias, our extensive human annotation study ($\kappa > 0.9$) confirms that these metrics are reliable proxies for evaluating functional equivalence. 

 \item \textbf{Execution reliability criteria.} Regarding the execution reliability metric ($M_{code}$), our pipeline strictly treats a non-zero exit code from the generated tool as a failure if the oracle executes successfully with a zero exit code. In practice, a non-zero exit code might not always signify a complete runtime crash, but rather a custom warning or specific state. However, such behavior violates common POSIX design principles~\cite{posix_standard} where successful operations should return zero. Enforcing this strict alignment provides the objective and standardized baseline for differential testing. 

 \item \textbf{Evaluation setting.} Our evaluation employs an unconstrained, zero-external-feedback setting where agents deliver the final repository in a single trajectory. While incorporating external test feedback (e.g., feeding error logs back to the agent) to enable multi-turn iterative debugging is a highly promising direction for Automated Program Repair, it falls beyond the scope of this paper, which focuses on evaluating the baseline 0-to-1 construction capability. Future iterations will aim to scale the dataset and explore these feedback-driven generation paradigms.

\end{itemize}

\section{Related Work}
\subsection{Agents for Software Engineering}
The application of LLMs in software engineering has evolved from static code completion to autonomous, environment-interacting agents~\cite{AutoAct, coder, Repairagent, autocoderover}. Early multi-agent frameworks, such as ChatDev~\cite{DBLP:conf/acl/QianLLCDL0CSCXL24} and MetaGPT~\cite{DBLP:conf/iclr/HongZCZCWZWYLZR24}, utilized role-playing and Standardized Operating Procedures (SOPs) to orchestrate software design via simulated collaboration. To enable real-world repository interaction, Agent-Computer Interfaces (ACIs) were introduced. Frameworks like SWE-agent~\cite{DBLP:conf/nips/YangJWLYNP24} and OpenHands~\cite{DBLP:conf/iclr/0001LSXTZPSLSTL25} allow LLMs to interact directly with command-line terminals, execute tests, and iteratively debug within isolated sandboxes. Recently, minimalist scaffolds like Mini-SWE-agent~\cite{mini-swe-agent} have demonstrated that state-of-the-art models require remarkably little scaffolding to achieve high success rates, relying primarily on their native reasoning capabilities.

\subsection{Software Generation and Agent Evaluation}
As agentic capabilities advance, evaluation methodologies have shifted to complex systems. Early benchmarks like HumanEval~\cite{DBLP:journals/corr/abs-2107-03374} and MBPP~\cite{DBLP:journals/corr/abs-2108-07732} focused on static, function-level synthesis but suffer from data contamination and a lack of repository context. Subsequent efforts addressed this through dynamic problem sourcing (LiveCodeBench~\cite{DBLP:conf/iclr/JainHGLYZWSSS25}) or complex API integration (BigCodeBench~\cite{DBLP:conf/iclr/ZhuoVCH0WYZHPB025}). For system-level tasks, SWE-bench~\cite{DBLP:conf/iclr/JimenezYWYPPN24} established the standard for software maintenance by evaluating patch generation for GitHub issues, while OSWorld~\cite{DBLP:conf/nips/XieZCLZCHCSLLXZ24} and Terminal-Bench~\cite{DBLP:journals/corr/abs-2601-11868} assess multi-step command execution in live operating systems.
However, evaluating zero-to-one software generation remains a critical challenge. Recent works like Commit0~\cite{DBLP:conf/iclr/ZhaoJLCCGR25} track an agent's ability to rebuild a repository commit-by-commit. While it scales to larger repositories, Commit0 provides pre-existing skeleton code and evaluates correctness through rigid static analysis and predefined unit tests. NL2Repo-Bench~\cite{DBLP:journals/corr/abs-2512-12730} explores repository-level generation from natural language, yet it also fundamentally relies on white-box unit tests. This rigid approach severely penalizes functionally correct but structurally diverse solutions. 
In parallel, Terminal-Bench~\cite{DBLP:journals/corr/abs-2601-11868} and OSWorld~\cite{DBLP:conf/nips/XieZCLZCHCSLLXZ24} evaluate multi-step bash command execution within live operating systems. While they test agent interactions with the terminal (e.g., navigating directories or installing packages), they neither focus on compiling a deliverable software entity from scratch nor strictly verify the resulting system-level file side-effects. In contrast, \dataset bridges these gaps by requiring full repository construction from an empty workspace and employing dynamic, black-box differential testing to rigorously validate both system side-effects and output equivalence.

\dataset~bridges these paradigms, extending black-box differential testing to the repository level to rigorously evaluate end-to-end software generation without imposing structural constraints.

\section{Conclusion}
% In this paper, we introduce \dataset, a benchmark for end-to-end evaluating LLM agents on software generation. Extensive evaluations reveal that the highest overall success rate remains below 44\%, indicating substantial room for future improvement in this challenging domain. Furthermore, our analysis uncovers several intriguing phenomena: agent performance exhibits a counterintuitive U-shaped trend with repository complexity, and higher token consumption does not equate to better task resolution. Finally, we observe that agents universally prefer monolithic designs over modular structures to simplify context management.
We introduce \dataset, a novel benchmark evaluating end-to-end software generation from scratch. Extensive evaluations reveal a maximum success rate of just 43.8\%, highlighting substantial room for improvement. Furthermore, our analysis uncovers key insights: performance exhibits a U-shaped trend with complexity, higher token consumption does not guarantee better results, and agents universally favor monolithic designs over modular ones to simplify context management.
Currently, we focus on CLIs. However, our core evaluation methodology of structure-agnostic, black-box differential testing is highly generalizable. In future work, we plan to extend this methodology to broader software domains, such as web services, API backends, and GUI applications, further advancing the assessment of fully autonomous software creation.

\section{Data Availability}
All scripts, datasets, and raw data are available at https://github.com/kinesiatricssxilm14/CLI-Tool-Bench.

\bibliographystyle{IEEEtran}
\bibliography{Citation}

\end{document}